\documentclass[11pt,twoside]{article}
\usepackage{asp2010}

\resetcounters

\begin{document}

\bibliographystyle{apj}

\title{Quasar Absorption Lines from Radiative Shocks: Implications for Multiphase Outflows and Feedback}
\author{Claude-Andr\'e Faucher-Gigu\`ere$^1$
\affil{$^1$Department of Astronomy and Theoretical Astrophysics Center, University of California, Berkeley, CA 94720-3411, USA.}}

\begin{abstract}
Photoionization modeling of certain low-ionization broad absorption lines in quasars implies very compact ($\Delta R \sim 0.01$ pc), galaxy-scale ($R\sim$ kpc) absorbers blueshifted by several 1000 km s$^{-1}$. 
While these are likely signatures of quasar outflows, the lifetimes of such compact absorbers are too short for them to be direct ejecta from a nuclear wind. 
Instead, I argue that the absorbing clouds must be transient and created $\emph{in situ}$. 
Following arguments detailed by Faucher-Gigu\`ere, Quataert, \& Murray (2011), I show that a model in which the cool absorbers form in radiative shocks arising when a quasar blast wave impacts an interstellar cloud along the line of sight successfully explains the key observed properties. 
Using this radiative shock model, the outflow kinetic luminosities for three luminous quasars are estimated to be $\dot{E}_{\rm k}\approx 2-5\%$ $L_{\rm AGN}$ (with corresponding momentum fluxes $\dot{P} \approx2-15 \, L_{\rm AGN}/c$), consistent with feedback models of the $M-\sigma$ relation. These energetics are similar to those recently inferred of molecular outflows in local ultra-luminous infrared galaxies and in post-starburt winds, suggesting that active galactic nuclei (AGN) are capable of driving such outflows. 
Radiative shocks probably affect the multiphase structure of outflows in a range of other systems, including narrower and higher-ionization quasar absorption lines, and compact intergalactic absorbers ejected by star formation and AGN activity.
\end{abstract}

\section{Introduction}
Recent observations have revealed compelling evidence for galaxy-scale outflows driven by AGN in ionized, neutral, and molecular gas \citep[e.g.,][]{2009ApJ...706..525M, 2011ApJ...733L..16S, 2011ApJ...729L..27R}. 
Such outflows are of particular interest in the context of correlations between galaxy-scale properties and those of the supermassive black holes (SMBH) they host, such as the $M-\sigma$ relation \citep[e.g.,][]{2002ApJ...574..740T}. 
Indeed, it is now a standard assumption that feedback by accreting SMBHs regulate their growth and produces these relationships \citep[e.g.,][]{1998A&A...331L...1S, 2003ApJ...595..614W, 2005Natur.433..604D}. 
It however remains an open question how the energy and/or momentum released by the SMBH couples to the surrounding galaxy. 
Observational constraints on AGN feedback are therefore extremely valuable.

\section{Radiative shock model for FeLoBALs}
Arguably the galaxy-scale quasar outflows with the most precisely measured physical properties are those traced by low-ionization broad absorption lines (BAL), in particular the FeLoBAL systems characterized by FeII$^{\star}$ absorption. 
The metastable transitions of FeII (and other low-ionization species, such as HeI, SiII, and NiII) are modeled to derive the parameters of the absorbing gas, including its number density and its ionization parameter $U_{\rm H} \propto L_{\rm AGN}/(R^{2} n_{\rm H})$. 
Since the AGN luminosity, $L_{\rm AGN}$, can also be measured, one can solve for the radius $R$ of the absorber from the SMBH. From these measurements, the energetics of the quasar outflows can in principle be derived.

The physical properties implied of FeLoBAL absorbers (tracing $T\sim10^{4}$ K gas), however, are surprising and demand a physical explanation before the derived energetics can be trusted. 
For three bright quasars ($L_{\rm AGN}=10^{46.7-47.7}$ erg s$^{-1}$), the photoionization models imply column and number densities $N_{\rm H}\sim10^{20}$ cm$^{-2}$ and $n_{\rm e}\sim10^{4}$ cm$^{-3}$, radii $R\sim1-3$ kpc, and blueshifts $v\sim1,000-5,000$ km s$^{-1}$ \citep[][]{2009ApJ...706..525M, 2010ApJ...709..611D, 2010ApJ...713...25B}. 
These properties give a characteristic size $\Delta R \equiv 1.2 N_{\rm H} / n_{\rm e} \sim10^{16}$ cm, corresponding to a dimensionless ratio $\Delta R/R \sim 10^{-5}$. 
These galaxy-scale absorbers therefore have masses as small as that of Jupiter!

\subsection{Transient nature}
Because of their high velocities, FeLoBAL absorbers were most likely accelerated by the bright central engine. While a common interpretation for BALs in general is that they arise in accretion disk winds \citep[][]{1995ApJ...451..498M, 2000ApJ...543..686P}, this hypothesis is ruled out for the FeLoBALs of interest due to the short lifetimes of the gas clouds \citep[][]{2011arXiv1108.0413F}. 
For the typical parameters outlined above, the FeLoBAL absorbers would in fact be destroyed by the Kelvin-Helmholtz instability or thermal evaporation by surrounding hot gas (necessary to pressure confine them) in just a few 1000 yr. 
On the other hand, the flow time $t_{\rm flow} \equiv R/v \approx 6\times10^{5}$ yr, for $R=3$ kpc and $v=5,000$ km s$^{-1}$. 
Thus, the compact absorbers must be transient clouds that formed approximately at the radius at which they are observed. 


\begin{figure} 
    \plotone{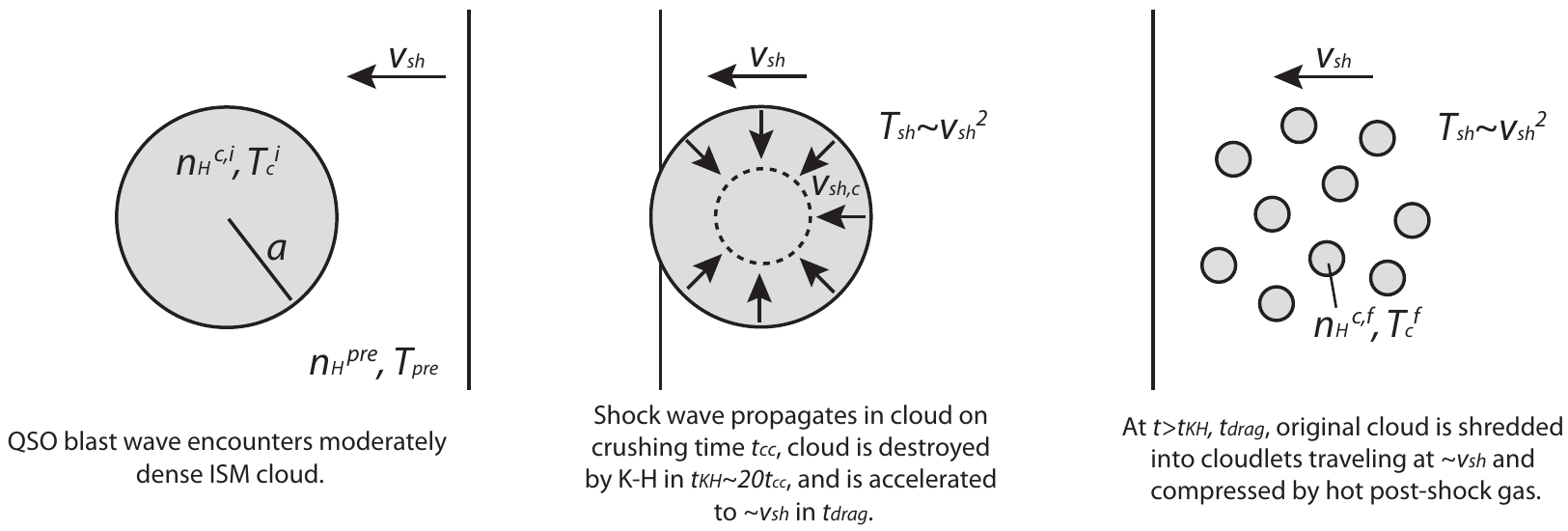}
    \caption{Schematic illustration of the formation of FeLoBALs in radiative cloud-crushing shocks \citep{2011arXiv1108.0413F}.}
    \label{cartoon} 
\end{figure}

\subsection{Formation in radiative shocks}
\cite{2011arXiv1108.0413F} introduced a specific mechanism for the \emph{in situ} formation compact cool absorbers, with properties in excellent agreement with observations of FeLoBALs. 
The model, illustrated schematically in Figure \ref{cartoon}, invokes the physics of cloud crushing \citep[e.g.,][]{1994ApJ...420..213K} to produce radiative shocks. 
Although the gas behind a quasar blast wave traveling at $v_{\rm sh}\sim5,000$ km s$^{-1}$ is too hot to cool in a flow time, the shocks propagating into an overdense cloud overcome by the blast wave are weaker, with a shock velocity reduced by a factor $\chi^{1/2}$, where $\chi \equiv n_{\rm H}^{\rm c}/n_{\rm H}^{\rm hot}$ is the density contrast of the cloud relative to the hot gas.

The two most basic requirements to produce absorbers consistent with FeLoBALs are: 1) the cloud-crushing shocks must be radiative, so that cool gas is left behind, and 2) the drag time, $t_{\rm drag}$, to accelerate the cloud by ram pressure from the hot gas must be less than the Kelvin-Helmholtz time, $t_{\rm KH}$, over which the cloud is destroyed. 
Remarkably, prescribing these basic requirements is sufficient to make a number of predictions, which quantitatively explain the observed FeLoBAL properties:
\begin{itemize}
\item The column density of the cool cloud is $N_{\rm H}^{\rm BAL} \gtrsim 10^{20}$ cm$^{-2}~\left(\frac{v}{\rm 5,000~km~s^{-1}} \right)^{4.2}$. 
\item The number density of the absorber is set by pressure equilibrium with the hot gas, $n_{\rm H}^{\rm BAL} \approx n_{\rm H}^{\rm hot} \left( \frac{T_{\rm hot}}{\rm 10^{4}~K} \right)$, where $T_{\rm hot}\approx6\times10^{8}~{\rm K}~\left( \frac{v}{\rm 5,000~km~s^{-1}} \right)^{2}$ 
is the post-shock temperature.\footnote{The shock velocity exceeds the velocity of the post-shock gas, which we identify with the velocity of the absorbing gas: $v_{\rm sh}=4v/3$.} The hot gas density $n_{\rm H}^{\rm hot} \approx 4 \bar{n}_{\rm H}^{\rm pre}(<R)$, where $\bar{n}_{\rm H}^{\rm pre}(<R)$ is the pre-shock mean density enclosed within $R$ and the factor of 4 arises from the jump conditions for a strong shock. 
$\bar{n}_{\rm H}^{\rm pre}(<R)$ itself can be estimated from the velocity of the swept up material and assuming that the AGN injects momentum at a rate $\sim L_{\rm AGN}/c$ in the galactic nucleus.  
For FeLoBAL parameters as above, $n_{\rm H}^{\rm BAL} \sim 3,000$ cm$^{-3}$. The predicted combination of $N_{\rm H}^{\rm BAL}$ and $n_{\rm H}^{\rm BAL}$ thus explains the small absorber sizes $\Delta R\sim10^{16}$ cm. 
\item The observational selection of low-ionization species, in particular FeII, selects a narrow range of ionization parameter $U_{\rm H} \propto L_{\rm AGN} / R^{2} n_{\rm H}$ and therefore defines a relationship between $L_{\rm AGN}$ and $R$. This explains why FeLoBAL absorbers in luminous quasars are found at large $R\sim$~kpc.
\item This picture, in which the absorbing gas is being shredded by instabilities, explains observations of multiple absorption components with approximately the same $R$ but different $v$ \citep[e.g.,][]{2010ApJ...713...25B}, since some cloudlets may only be partially accelerated. 
The supra-thermal line widths measured may also result from a combination of blended components and velocity shear. \item The fact that FeLoBALs arise along overdense sight lines is in agreement with their host quasars being redder than average \citep[e.g.,][]{2000ApJ...538...72B}.
\end{itemize}

\begin{table}
\label{tab:feedback}
\centering
\caption{Galaxy-scale mass outflow rates, kinetic luminosities, and momentum fluxes inferred for three bright FeLoBAL quasars}
\begin{tabular}{|ccccc|}
\hline\hline
QSO                                & $\dot{M}_{\rm hot}(\Omega_{\rm hot}=1)$ & $\dot{E}_{\rm k}^{\rm hot}$ & $\dot{P}^{\rm hot}$ & $\dot{E}_{\rm k}^{\rm QSO}$ \\
                                      & M$_{\odot}$ yr$^{-1}$ & $\%L_{\rm AGN}$ & $L_{\rm AGN}/c$ & $\%L_{\rm AGN}$ \\
\hline
SDSS J0838+2955 & 1,000 & 2.2 & 2.7 & 3.5 \\ 
SDSS J0318-0600 & 1,100 & 1.2 & 1.8 & 2.0 \\ 
QSO 2359-1241 & 2,400 & 3.1 & 13.4 & 5.0 \\ 
\hline
\end{tabular}

Values for the hot flow obtained using our radiative shock model are labeled `hot.' 
$\dot{E}_{\rm k}^{\rm QSO}$ gives the kinetic luminosity of the small-scale quasar wind inferred assuming a self-similar adiabatic wind solution (for details, see Faucher-Gigu\`ere et al. 2011). 
Based on data from \cite{2009ApJ...706..525M}, \cite{2010ApJ...709..611D}, and \cite{2010ApJ...713...25B}.
\end{table}

\subsection{Implications of multiphase structure for outflow energetics}
\label{sec:energetics}
A key implication of our radiative shock model is that the FeLoBAL absorbers are only tracers of an underlying multiphase outflow, which can be much more massive. 
On the other hand, it is commonly assumed for the purpose of deriving energetics that the outflowing material is confined to a cold, thin shell of covering factor $\Omega$, column density $N_{\rm H}^{\rm BAL}$, and velocity $v$.  
Since the mass of the shell $M_{\rm shell} = 4 \pi R^{2} N_{\rm H}^{\rm BAL} m_{\rm p} \Omega / X$ (where $X$ is the hydrogen mass fraction), the mass outflow rate\footnote{This expression assumes that $N_{\rm H}^{\rm BAL}$ is constant in time; the dimensionless pre-factor should be regarded as uncertain at the order unity level.} 
\begin{equation}
\label{Mdot shell}
\dot{M}_{\rm shell} = \frac{8 \pi R N_{\rm H}^{\rm BAL} m_{\rm p} v} {X} \Omega
\end{equation}
\citep[e.g.,][]{2010IAUS..267..350A}. 
The corresponding kinetic luminosity and momentum flux are then 
\begin{equation}
\label{Edot k}
\dot{E}_{\rm k} = \frac{1}{2} \dot{M}_{\rm shell} v^{2};~~~~~\dot{P} = \dot{M}_{\rm shell} v.
\end{equation}

In our picture, most of the outflowing mass is instead carried by a hot flow. More accurate outflow energetics are therefore obtained by replacing $N_{\rm H}^{\rm BAL} \to N_{\rm H}^{\rm hot}$ in the above expressions. 
The properties of the hot flow are not directly measured, but can be inferred in the context of our model. Specifically, we assume that the FeLoBAL gas is fully accelerated by the hot gas, $v \approx v_{\rm hot}$, and in pressure equilibrium with it, $n_{\rm H}^{\rm hot} \approx n_{\rm H}^{\rm BAL} \left( \frac{\rm 10^{4}~K}{T_{\rm hot}} \right)$. 
The column density of the hot gas can be estimated as $N_{\rm H}^{\rm hot} \approx \bar{n}_{\rm H}^{\rm pre}(<R) R \approx n_{\rm H}^{\rm hot} R/4$. 
Under these assumptions,
\begin{equation}
N_{\rm H}^{\rm hot} \approx \frac{1}{4} N_{\rm H}^{\rm BAL} \left( \frac{R}{\Delta R} \right) \left( \frac{\rm 10^{4}~K}{T_{\rm hot}} \right),
\end{equation}
where  $\Delta R$ is the thickness of the FeLoBAL absorber (as before). 
Written this way, all the quantities necessary to evaluate equations (\ref{Mdot shell}, \ref{Edot k}) are provided by photoionization modeling of the FeLoBAL, except $T_{\rm hot}$ which is derived from $v$, and $\Omega_{\rm hot}$. 

In Table \ref{tab:feedback}, we give the mass outflow rates, kinetic luminosities, and momentum fluxes implied by the radiative shock model for three bright FeLoBAL quasars in the literature (for details, see Faucher-Gigu\`ere et al. 2011). For these estimates, we fiducially adopt a covering factor $\Omega_{\rm hot}=1$ for the hot gas. 
While covering factors $\approx0.2$ motivated by the general incidence of BALs in quasars are often assumed, this assumption is not justified in our model. 
Indeed, if most BALs arise in accretion disk winds close the central SMBH, then the covering factor of the galaxy-scale quasar outflows probed by FeLoBALs can be very different. 
In the absence of strong anisotropy in the ambient medium, the outflow in facts tends to become quasi-spherical as it expands. 

Interestingly, the kinetic luminosities of the quasar outflows in Table \ref{tab:feedback} are $\approx2-5$\% $L_{\rm AGN}$ (with momentum fluxes $\approx 2-15~L_{\rm AGN}/c$), in good agreement with feedback models of the $M-\sigma$ relation \citep[][]{2005Natur.433..604D, debuhr11}. 
Similar energetics have also been recently inferred of molecular outflows in local ultra-luminous infrared galaxies \citep[][]{2011ApJ...733L..16S} and post-starburst galaxies \citep[][]{2007ApJ...663L..77T}. 
This suggests that AGN can also drive the outflows in those systems, which are physically connected to luminous quasars in merger-driven scenarios \citep[e.g.,][]{2008ApJS..175..356H}. 
It is worth noting that the energetics inferred in the context of our multiphase, radiative shock model exceed the values implied by the cold thin shell approximation by orders of magnitude in certain cases. 

\section{Applications to other systems}
Although we developed the radiative shock model to explain the extraordinary properties of FeLoBALs, similar physical processes are likely important in determining the multiphase structure of outflows in a broad range of systems. 
These potentially include some higher-ionization and narrower quasar absorption lines \citep[e.g.,][]{2011MNRAS.410.1957H}. 
Many intergalactic absorbers, probably associated with star-formation powered galactic winds, are also inferred to be compact and transient \citep[e.g.,][]{2006ApJ...637..648S, 2007MNRAS.379.1169S}. 
Finally, it was recently shown that the hot gas mass in a post-starburst wind exceeds that of the cool gas by a factor $10-150$ \citep[][]{2011arXiv1111.3982T}. 

\acknowledgements I thank Eliot Quataert and Norm Murray for collaboration on the subject of this proceeding.

\bibliography{references}

\begin{thebibliography}{}
\expandafter\ifx\csname natexlab\endcsname\relax\def\natexlab#1{#1}\fi
\expandafter\ifx\csname url\endcsname\relax
  \def\url#1{\texttt{#1}}\fi
\expandafter\ifx\csname urlprefix\endcsname\relax\def\urlprefix{URL }\fi
\providecommand{\eprint}[2][]{\url{#2}}

\bibitem[{{Arav}(2010)}]{2010IAUS..267..350A}
{Arav}, N. 2010, in IAU Symposium, vol. 267 of IAU Symposium, 350

\bibitem[{{Bautista et al.}(2010)}]{2010ApJ...713...25B}
{Bautista et al.} 2010, \apj, 713, 25. \eprint{1003.0970}

\bibitem[{{Becker et al.}(2000)}]{2000ApJ...538...72B}
{Becker et al.} 2000, \apj, 538, 72. \eprint{arXiv:astro-ph/0002470}

\bibitem[{{DeBuhr} et~al.(2011){DeBuhr}, {Quataert}, \& {Ma}}]{debuhr11}
{DeBuhr}, J., {Quataert}, E., \& {Ma}, C.-P. 2011, ArXiv e-prints.
  \eprint{1107.5579}

\bibitem[{{Di Matteo} et~al.(2005){Di Matteo}, {Springel}, \&
  {Hernquist}}]{2005Natur.433..604D}
{Di Matteo}, T., {Springel}, V., \& {Hernquist}, L. 2005, \nat, 433, 604.
  \eprint{arXiv:astro-ph/0502199}

\bibitem[{{Dunn et al.}(2010)}]{2010ApJ...709..611D}
{Dunn et al.} 2010, \apj, 709, 611. \eprint{0911.3896}

\bibitem[{{Faucher-Gigu\`ere} et~al.(2011){Faucher-Gigu\`ere}, {Quataert}, \&
  {Murray}}]{2011arXiv1108.0413F}
{Faucher-Gigu\`ere}, C.-A., {Quataert}, E., \& {Murray}, N. 2011, ArXiv
  e-prints. \eprint{1108.0413}

\bibitem[{{Hamann et al.}(2011)}]{2011MNRAS.410.1957H}
{Hamann et al.} 2011, \mnras, 410, 1957. \eprint{1008.3728}

\bibitem[{{Hopkins} et~al.(2008){Hopkins}, {Hernquist}, {Cox}, \& {Kere{\v
  s}}}]{2008ApJS..175..356H}
{Hopkins}, P.~F., {Hernquist}, L., {Cox}, T.~J., \& {Kere{\v s}}, D. 2008,
  \apjs, 175, 356. \eprint{0706.1243}

\bibitem[{{Klein} et~al.(1994){Klein}, {McKee}, \&
  {Colella}}]{1994ApJ...420..213K}
{Klein}, R.~I., {McKee}, C.~F., \& {Colella}, P. 1994, \apj, 420, 213

\bibitem[{{Moe} et~al.(2009){Moe}, {Arav}, {Bautista}, \&
  {Korista}}]{2009ApJ...706..525M}
{Moe}, M., {Arav}, N., {Bautista}, M.~A., \& {Korista}, K.~T. 2009, \apj, 706,
  525. \eprint{0911.3332}

\bibitem[{{Murray} et~al.(1995){Murray}, {Chiang}, {Grossman}, \&
  {Voit}}]{1995ApJ...451..498M}
{Murray}, N., {Chiang}, J., {Grossman}, S.~A., \& {Voit}, G.~M. 1995, \apj,
  451, 498

\bibitem[{{Proga} et~al.(2000){Proga}, {Stone}, \&
  {Kallman}}]{2000ApJ...543..686P}
{Proga}, D., {Stone}, J.~M., \& {Kallman}, T.~R. 2000, \apj, 543, 686.
  \eprint{arXiv:astro-ph/0005315}

\bibitem[{{Rupke} \& {Veilleux}(2011)}]{2011ApJ...729L..27R}
{Rupke}, D.~S.~N., \& {Veilleux}, S. 2011, \apjl, 729, L27. \eprint{1102.4349}

\bibitem[{{Schaye et al.}(2007)}]{2007MNRAS.379.1169S}
{Schaye et al.} 2007, \mnras, 379, 1169. \eprint{arXiv:astro-ph/0701761}

\bibitem[{{Silk} \& {Rees}(1998)}]{1998A&A...331L...1S}
{Silk}, J., \& {Rees}, M.~J. 1998, \aap, 331, L1.
  \eprint{arXiv:astro-ph/9801013}

\bibitem[{{Simcoe et al.}(2006)}]{2006ApJ...637..648S}
{Simcoe et al.} 2006, \apj, 637, 648. \eprint{arXiv:astro-ph/0508116}

\bibitem[{{Sturm et al.}(2011)}]{2011ApJ...733L..16S}
{Sturm et al.} 2011, \apjl, 733, L16+. \eprint{1105.1731}

\bibitem[{{Tremaine et al.}(2002)}]{2002ApJ...574..740T}
{Tremaine et al.} 2002, \apj, 574, 740. \eprint{arXiv:astro-ph/0203468}

\bibitem[{{Tremonti} et~al.(2007){Tremonti}, {Moustakas}, \&
  {Diamond-Stanic}}]{2007ApJ...663L..77T}
{Tremonti}, C.~A., {Moustakas}, J., \& {Diamond-Stanic}, A.~M. 2007, \apjl,
  663, L77. \eprint{0706.0527}

\bibitem[{{Tripp et al.}(2011)}]{2011arXiv1111.3982T}
{Tripp et al.} 2011, ArXiv e-prints. \eprint{1111.3982}

\bibitem[{{Wyithe} \& {Loeb}(2003)}]{2003ApJ...595..614W}
{Wyithe}, J.~S.~B., \& {Loeb}, A. 2003, \apj, 595, 614.
  \eprint{arXiv:astro-ph/0304156}

\end{thebibliography}

\end{document}